%% file: template.tex
\documentclass{vgtc}                          

\ifpdf
  \pdfoutput=1\relax                   
  \usepackage{graphicx}                
  \DeclareGraphicsExtensions{.pdf,.png,.jpg,.jpeg} 
\else
  \usepackage{graphicx}                
  \DeclareGraphicsExtensions{.eps}     
\fi%

\graphicspath{{figures/}{pictures/}{images/}{./}} 

\usepackage{microtype}                 
\PassOptionsToPackage{warn}{textcomp}  
\usepackage{textcomp}                  
\usepackage{mathptmx}                  
\usepackage{times}                     
\usepackage{cite}                      
\usepackage{tabu}                      
\usepackage{booktabs}                  

\onlineid{0}

\vgtccategory{Research}

\vgtcinsertpkg

\newcommand{\app}{\emph{Sprite}}

\title{Evaluating how interactive visualizations can assist in finding samples where and how computer vision models make mistakes}

\author{Hayeong Song\thanks{e-mail: hsong300@gatech.edu}\\ %
        \scriptsize Georgia Institute of Technology %
\and Gonzalo Ramos\thanks{e-mail: goramos@microsoft.com}\\ %
     \scriptsize Microsoft Research  %
\and Peter Bodik\thanks{e-mail: peterb@microsoft.com}\\ %
     \scriptsize Microsoft Research}

\teaser{
  \centering
  \includegraphics[width=0.8\linewidth]{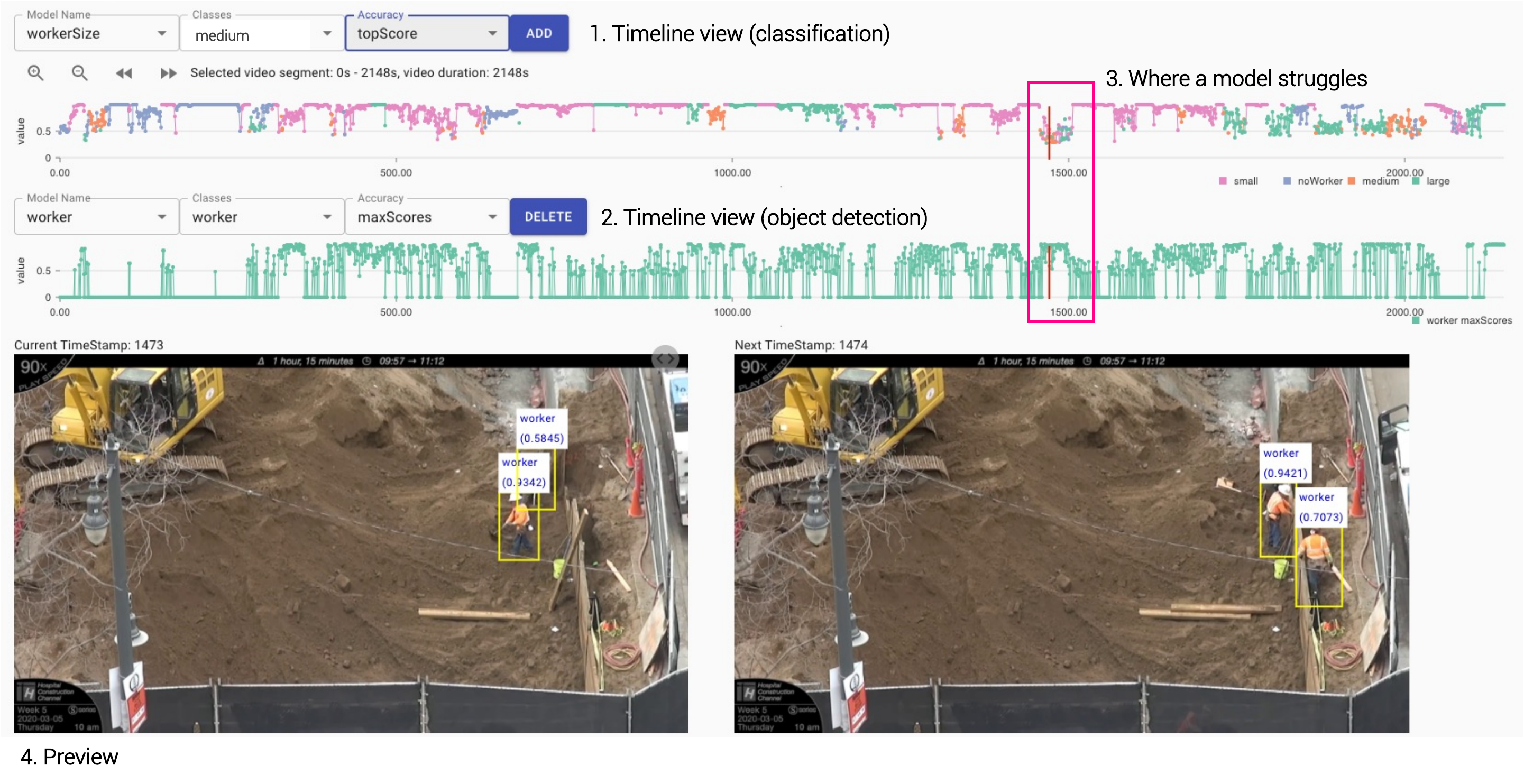}
  \caption{\app's timeline view for a model that classifies images based on the size (small, medium, and large) of the worker in view. A user can utilize this view in the following way:
  \textbf{1}. The user can display the worker size classification prediction results in a timeline view, which shows prediction results over time (when a score is high, the model is confident about its prediction). Color coding shows the predicted size classification (label) of the worker for each video frame. 
  \textbf{2}. The user can add another timeline view that shows prediction results for worker object detection model. The preview images include bounding boxes where the model detects a worker. 
  \textbf{3}. The user can find an area in the timeline (pink rectangle) where the model seems to struggle. For example, the pink rectangle in the top portion of the preview above, the model struggled to classify a sequence of images with similar traits over a short period of time. The view differentiates its borderline prediction scores based on predicted labels, such as small (pink) and large (green). 
\textbf{4.} After reviewing a sequence of images, the user finds out that the model often misclassifies images of medium-sized workers. The user is then able to select meaningful new training images (with medium-sized workers) to improve the model's performance.}
\label{fig:time-series}
}

\abstract{Creating Computer Vision (CV) models remains a complex practice, despite their ubiquity.  Access to data, the requirement for ML expertise, and model opacity are just a few points of complexity that limit the ability of end-users to build, inspect, and improve these models. Interactive ML perspectives have helped address some of these issues by considering a teacher in the loop where planning, teaching, and evaluating tasks take place. We present and evaluate two interactive visualizations in the context of Sprite, a system for creating CV classification and detection models for images originating from videos. We study how these visualizations help Sprite's users identify (evaluate) and select (plan) images where a model is struggling and can lead to improved performance, compared to a baseline condition where users used a query language. We found that users who had used the visualizations found more images across a wider set of potential types of model errors.} 

\CCScatlist{
  \CCScatTwelve{Human-centered computing}{Visu\-al\-iza\-tion}{Human computer interaction (HCI)}{HCI design and evaluation methods}{---Usability testing};
  \CCScatTwelve{Human-centered computing}{Visu\-al\-iza\-tion}{Visualization application domains}{Visual analytics}
}

\begin{document}




\maketitle

\input{sections/01-introduction}
\input{sections/02-related_work}

\input{sections/03-system}

\input{sections/04-study}

\input{sections/05-results}

\input{sections/06-discussion}

\input{sections/07-conclusion}

\bibliographystyle{abbrv-doi}

\bibliography{template}
\end{document}

%% file: sections/01-introduction.tex
\section{Introduction}

Research in the field of interactive Machine Learning (ML) and teaching \cite{dudley2018review, ramos2020imt} has made progress in reducing the complexity of creating ML models by having a human in the loop~\cite{amershi2014power, dudley2018review, fails2003interactive, holzinger2019interactive}. This interactive loop can be characterized by curriculum building, knowledge transfer, and model evaluation stages \cite{ramos2020imt}. While defining these teaching tasks is in principle straightforward, it presents challenges. In the case of curriculum building, the existence of unknowns (e.g., examples a model has not seen or features an ML model is blind to) and the presence of bias are obstacles that can get in the way of the task~\cite{suh2019anchorviz}. The transfer of knowledge can be limited to the inputs a learning algorithm permits, and challenges still persist in allowing teachers in the loop to express rich knowledge~\cite{sultanum2020teaching, ng2020understanding}.
Model evaluation is an activity that has a direct relationship with topics in the area of explainable ML (XAI)~\cite{ribeiro2016should, adadi2018peeking}. In the context of an interactive ML model-building process, the system can only provide feedback and confidence about what it knows.

This paper describes an augmentation to \app~that allows users to train CV models that get applied to images originating from video footage. We focus on the planning and evaluation stages, seeking ways not only to support subject domain experts (people with knowledge germane to the task the model is performing and those who do not necessarily have ML expertise) in efficiently identifying challenging documents (images) in which the model is not performing well but also to identify the potential reasons why. Our enhancement to \app~ supports users facing these challenges by improving the interactive machine teaching process so that a user and the learning system collaborate through interactive visualizations to find images that should be labeled (and added to the training dataset) to improve the CV model. We propose two interactive visualizations, \emph{timeline view} and \emph{scatterplot view}, in the context of \app, a system to build CV models from video footage.

We conducted a usability study to assess the effects of interactive visualizations. This study looked into how our views--the timeline and scatterplot views--help participants identify (evaluate) and select (plan) images about which a model is struggling to make correct predictions. We tested two conditions: baseline (\app~without our augmentation) and visualization (\app~with our augmentation). With this assessment, our work offers the following contributions. First, we’ve designed interactive visualizations and have studied how they can help users (who are training CV models) more efficiently sample images that contain diverse prediction errors in the context of \app~(see Section~\ref{section:system}). These visualizations support users in browsing and finding candidate images and in assessing and contrasting the prediction behavior of one or more models (e.g., classification and object-detection models). Second, we present the results of a study that shows that participants using these interactive visualizations were more efficient in finding candidate images that stumped the model when using a \emph{baseline} condition. Finally, we provide insights, discuss design implications, and suggest future directions to support subject-domain experts during interactive machine teaching of CV models.

 



%% file: sections/02-related_work.tex
\section{Related Work}
\label{section: related work}

\subsection{Interactive machine learning and teaching} 
There is a great body of research in the area of interactive machine learning (IML) where a person in the loop ``iteratively builds and refines a mathematical model to describe a concept through iterative cycles of input and review'' \cite{dudley2018review, dang2022ganslider}. Systems like the Wekinator \cite{fiebrink2010wekinator} and NorthStar \cite{kraska2018northstar} are but just a small example of IML experiences making ML accessible for end-users. There are different perspectives one can take when thinking about the skills, roles, and goals of people in the loop in interactive machine teaching (IMT) systems. IMT is an opinionated perspective on IML where people in the loop take on the role of a teacher of a particular concept and interact with the learning system in a ``teacher-student-type of interaction''. In the IMT perspective, people (teachers) in the loop are engaged in planning and updating the teaching curriculum, explaining knowledge pertinent to the subject domain, and reviewing the learner's progress while integrating the given knowledge. In our work, we explore IML as the general framework for helping people create CV models and use IMT's perspective to unpack the activities people partake in while training the model. Our work focuses on the planning and reviewing stages. For a comprehensive list of articles on the topic of IML and IMT, we refer readers to these works \cite{dudley2018review, amershi2014power,wu2022survey,simard2017machine, ramos2020imt}.

\subsection{Building CV models interactively} 
Systems and prototypes like Lobe.ai~\footnote{https://lobe.ai/} and Google's teachable machines~\footnote{https://teachablemachine.withgoogle.com/} provide experiences that hide the complexity of directly interacting with a learning algorithm design and parameters and instead, present interfaces where people in the loop only need their subject-domain expertise to teach to the system. While lowering the barrier of entry for end-users, these tools only focus on part of the interactive machine teaching cycle, limiting evaluation and leaving behind tasks such as planning. Current advances in transfer learning make it possible for people to create ML models with less effort and labels by leveraging pre-existing large models \cite{oquab2014learning,yosinski2014transferable}. This makes it possible for services like Azure's Custom Vision~\cite{customvision2019} to provide people with experiences where they only need to provide a fraction of the labels that otherwise building a robust CV model would take.
Our work leverages this type of learning in its own experience, providing people with different opportunities to review a model and select the most relevant labeled data to present next.


%% file: sections/03-system.tex
\section{The \app~system}
\label{section:system}

\begin{figure*}[h]
\centering
\includegraphics[width=0.75\textwidth]{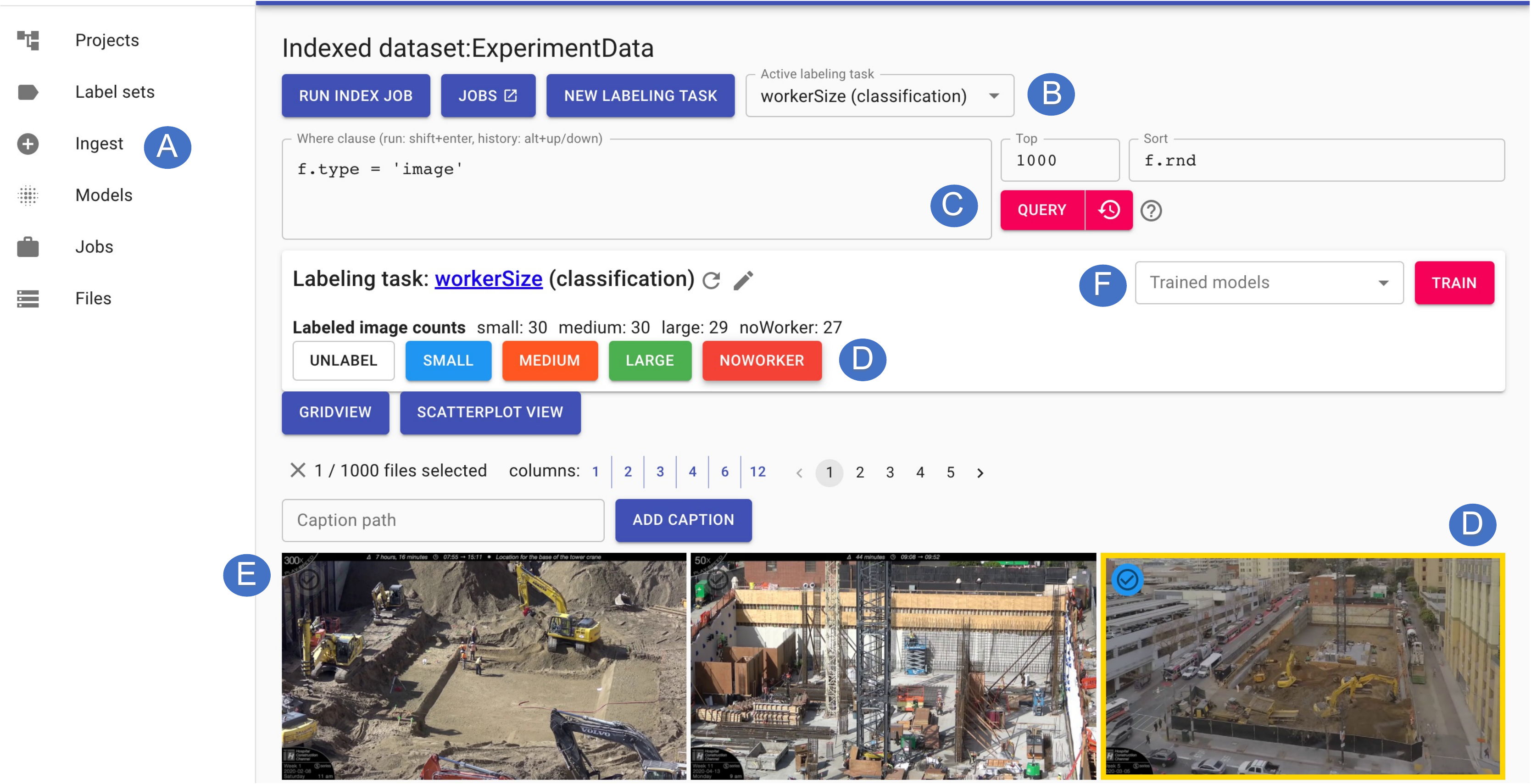}
\caption{General UI of a \app\ system. A-The system's menu shows where people can ingest (upload) videos into the system. B-Selector for the current model being worked/trained on. C-Main sampling query area to retrieve from the set of decoded images. D-Labeling selector that indicates the current classes of a selected classification model, and that is used to indicate what label to assign to a selected image. E-Grid of retrieved images from a query. F-Control to initiate a training operation and check prediction accuracy.}
\label{fig:UI}
\end{figure*}

\app~implements an end-to-end interactive ML workflow in the context of creating CV models using images extracted from video footage. A system like this can be used to produce predictive models to help monitor video sources such as security feeds or game streaming for content of interest. While \app~follows the standard structure found in many interactive machine learning and teaching systems, \emph{it is not the main focus of the article;} We write about it only to provide a fuller picture of our work. Our contributions do not focus on \app~itself but, rather, on studying the effects of using interactive visualizations on discovering diverse examples of images about which models made prediction errors. In this section, we discuss the design goals and relevant implementation details of our contribution. 

\subsection{Design Goals}
\label{section:desgin-goals}

\begin{enumerate}
\itemsep-0.2em
  \item[G1.] \textbf{Allow easy discovery of diverse prediction error patterns.} Improving sampling means being able to efficiently find diverse samples where a model struggles (i.e., variation of object appearances~\cite{gou2020vatld}). 
  \item[G2.] \textbf{Support leveraging and comparing existing CV models to find images with prediction errors.} People are naturally attuned to react to both cognitive and visual dissonances or differences. 
  \item[G3.] \textbf{Help subject-domain experts discover prediction patterns and errors across a video's timeline.} Predictions over data that has a temporal component (e.g., temporal stability~\cite{ren2016squares,hohman2020understanding}) can reveal insightful patterns about a model's performance, that cannot be observed by inspecting individual predictions in isolation.
  \item[G4.] \textbf{Increase people's access to ML building tools by reducing the need for ML literacy.} We aspire to create experiences where subject domain experts can build CV models in efficient ways. Of the many ways to achieve this goal, we seek to leverage the direct agencies that interactive visualization can provide~\cite{choo2018visual,liu2017towards,zhang2018visual}.
\end{enumerate}

\textbf{General System Architecture. }\app~is implemented as a web app using backend services that handle tasks, such as model training and applying the model(s) to video frames. \app~performs transfer learning using \emph{Resnet50}~\cite{he2015deep} for classification tasks, and \emph{YOLOv4-tiny}~\cite{bochkovskiy2020yolov4} for object detection tasks. Model evaluation is performed on a prediction cluster that uses multiple GPUs to process images with high throughput. 
After performing an inference on a specific image, the image metadata in the database is updated with the inference output. All processed video frames are indexed in Azure Cosmos DB store, where users can retrieve relevant images using SQL queries~\cite{cosmosdb-docs} at interactive speeds.

\textbf{General System Flow \& General UI. }In \app, users are engaged in the following workflow when they train and evaluate CV models: 1) upload videos (see Figure~\ref{fig:UI} (a)), 2) decode video to frames (images), 3) sample decoded images that meet a target concept, 4) label samples images (see Figure~\ref{fig:UI} (d)), 5) train CV models, and 6) evaluate model performance and predictions. To check prediction results, users can see prediction results for the in-process CV model. This paper focuses on stages 3 and 6, where people sample and evaluate decoded video frames to improve the next CV model. We specifically focus on these two phases because they are critical phases for debugging (evaluating) and improving (planning) a CV model in an interactive machine teaching loop~\cite{ramos2020imt}.

\subsection{Interactive Visualizations}

We designed the interactive views (design process) based on 1) a review of ML exploration systems, 2) characterization and extracted design alternatives from literature, and 3) discussions on trade-offs on design alternatives. Our design was an iterative process, which involved researchers discussing design alternatives (i.e., confusion matrix~\cite{chen2016diagnostic}), prototyping, and informal usability tests with our target users (e.g., novices in ML to knowledgeable in ML) to refine our design. Following our design goals, we support two main interactive visualizations, which are \textit{timeline view} and \textit{scatterplot view} (see Section~\ref{section:desgin-goals} for design goals). We support these views to help people in the loop to sample a set of images (labeled \& unlabeled) where a CV model makes prediction errors (G4). This way, users can interact with visualization to explore different aspects of the data and model, which can allow users to formulate hypotheses on why CV models make mistakes and gain insight into the IMT process. We choose these views based on our design goals, which can help users review images on a large scale and help them understand local and global prediction patterns at a glance. 


\textbf{Timeline view.} Our first design follows our goals to help people find images with potential errors by leveraging and comparing existing CV models (G2) and to discover prediction patterns/errors across a video's timeline (G3). A timeline view provides a holistic view of prediction patterns for decoded video frames over time. This allows people to quickly find suspicious patterns such as spikes and dips globally and locally by relying on the temporal consistency in a video (see Figure ~\ref{fig:time-series} (1) \& (2)). Users can add and stack up to three graphs in a configuration that allows them to check prediction results across different CV models. This feature can help people troubleshoot CV models by comparing multiple models (e.g., object detection and classification models). For example, if a classification model predicted an image to have a worker, but a detection model did not detect any worker, then one of the models is wrong. This disparity between the prediction results of CV models is a strong indicator that an image is worth inspecting. For \textbf{\textit{classification}}, a timeline view displays results with color coding (each prediction class is represented with a different color). This view can help view global/local patterns over time, as users can see prediction results in groups (see Figure~\ref{fig:time-series} (1)). For a classification model, \texttt{scores} represent the prediction scores for individual classes, \texttt{topScore} is the highest score, and \texttt{topClass} is the class with the highest score (scores range from 0 - 1). Note that a prediction with a low \texttt{topScore} (e.g., 0.4 ${<}$ score ${<}$ 0.7) is a sign that the model is not confident in the prediction. For \textbf{\textit{detection}}, the system annotates preview image(s) with bounding boxes and shows scores of what the detection model detected (see Figure~\ref{fig:time-series} (4)). For object detectors, \texttt{classes} is a list of detected classes in the image, \texttt{counts} is the number of detected objects of each class, and \texttt{maxScore} is the maximum score across all detections of a certain class.

\begin{figure*}[h!]
\centering
\includegraphics[width=0.8\textwidth]{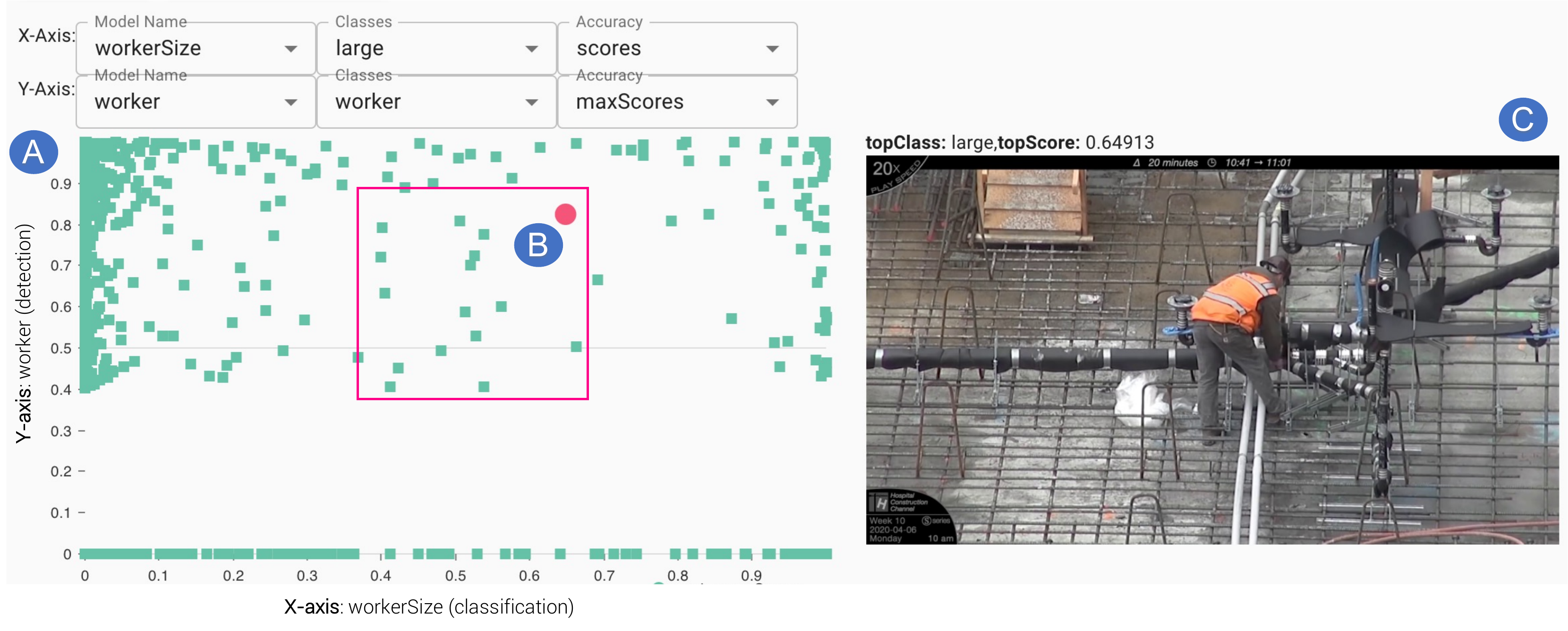}
\caption{A - Scatterplot view used to inspect the predictions of two CV models: \emph{workerSize (classification)} and \emph{worker (detection)}. The X-axis displays worker size classifier prediction results. The Y-axis displays prediction results for a worker detection model. B - The selected data point (the red circle) represents an image where the two models seem to agree. C - The image corresponding to the selected data point is shown along with a prediction label and score for the model currently being evaluated. 
To troubleshoot a classification model, a user can check data points corresponding to borderline prediction scores (pink area, e.g. 0.4 ${<}$ score ${<}$ 0.7). Then the user can check samples where a worker was detected (high worker detection scores) and borderline scores for a classification model, as the models seem to struggle (i.e., noisy background). 
This type of visualization can be helpful to quickly retrieve sub-samples of images that can help improve a model, by assisting users to compare prediction results across semantically related CV models.}
\label{fig:scatterplot}
\end{figure*}

\textbf{Scatterplot View.} Our follow-up design stems from our goals to allow people to discover diverse error patterns easily (G1) and help people easily leverage and compare existing CV models (G2). In the context of our system, we call this design the scatterplot view (see Figure~\ref{fig:scatterplot}). This view can help users find prediction errors faster because it shows similar images in clusters (based on people-specified metrics) and helps people find corner cases, where two CV models' prediction results do not agree on. To check prediction, users can select what they want the x- and y-axis to represent to compare prediction results for two models. Users can select a model, class name (if applicable), and accuracy through the corresponding drop-down menus. 
For example, users can look for outliers to see if there is a disparity between the two models' prediction results. An example of an outlier or irregularity is an image that has been classified as \texttt{noWorker} by a classification model, but where a detection model detected a worker. Then they retrieve an image of interest to inspect, by clicking on a data point on the plot (see Figure~\ref{fig:scatterplot}).

%% file: sections/04-study.tex
\section{Study Design}

We wanted to assess how our proposed interactive visualizations can help users find a diverse set of examples where the model they are building makes prediction errors. In designing our study, we considered allowing participants to complete a few model training cycles to assess the success of the provided support but decided to instead observe the diversity of the examples that participants selected. This is because adding full training cycles would have significantly extended the duration of the study, compromising both human factors such as comfort and fatigue. Our study then consisted of a between-subjects experiment (\emph{baseline} and \emph{visualization}) to understand how our visualizations support people to be more efficient in finding meaningful video frames (images) to improve a CV model. In the \emph{baseline} condition, participants could use a query language to sample images and participants in \emph{visualization} condition could use both query languages and visualization. We randomly divided the 20 participants into two groups of 10 each. After the study's main task, participants filled out a survey where we collected qualitative information about their experience as well as quantitative scores to measure usability~\cite{brooke1996quick} and subjective workload~\cite{hart2006nasa}) based on a 5-point Likert survey.

\textbf{Participants.} We recruited 20 participants (13 male, 7 female), all of whom work for a large technology company. Participants came from diverse backgrounds that included design, research, and project management. Only 35\% of the participants reported that they used ML for CV tasks. Participants with this type of background fit the profile of subject-domain experts, that have various levels of ML and CV model training experience. The study took about 90 minutes to complete. We compensated participants with a \$50 gift card.


\textbf{Dataset and Models.} The dataset we used was about a hospital construction site from a Youtube Channel~\cite{youtube} (6 videos, ~40 minutes long) were uploaded in \app. For the user study, we presented participants with a system where a number of CV models were in the process of being trained. The set-up consisted of participants taking over the process of identifying sample images where in-training CV models struggled, which could then be used to fuel a subsequent labeling and training stage. The in-progress CV models we included in the study consisted of two classification models and a detection model. These \textbf{classifications} models were, first \emph{view classifier} classifies images depending on the camera zoom level (labels: \texttt{closeup}, \texttt{medium}, and \texttt{full}) in the scene. Second, \emph{workerSize classifier}, classifies images based on worker height compared to image size (labels: \texttt{small}, \texttt{medium}, \texttt{large}, and \texttt{noWorker}). A \textbf{detection} model detects the presence and location of construction workers in an image. All of the models had above 90\% accuracy.  

\textbf{Procedure.} We conducted the studies using the Teams teleconferencing platform. Participants used their laptops to access a link to the particular experimental condition we assigned them to. The general procedure for the experiment was the same. We first asked participants to fill out a demographic survey. Then participants watched a 13-minute tutorial video, which explained the main features of \app, such as how to use queries, read prediction results, and demonstrate examples where the model made prediction errors.

\textbf{User Task.} After onboarding and practicing, we asked participants to find images, where a CV model makes prediction errors for a classification model and a detection model. We gave 20 minutes to complete each task. To retrieve images, sometimes participants had to write queries. To allow participants to explore images freely, we provided example queries participants could use and the first author (researcher) facilitated the study and served as \textit{wizard}~\cite{steinfeld2009oz} to alleviate the need for the participants to know query language and to make the contrast between conditions as fair as possible. For the \emph{classification} model, we asked participants to find images, where the \emph{workerSize} model makes a prediction error. For example, the model might have classified images with a \texttt{large} worker to a \texttt{medium} worker. For the \emph{detection model}, we asked participants to find images where the \emph{worker} model makes prediction errors. The model might have also missed worker objects. For example, an image might have three workers, but a model might have only detected one of them. In the \textbf{\textit{baseline}} condition, we asked participants to find an image where a model made prediction errors using querying language and general UI (Figure~\ref{fig:UI}). For the \textbf{\textit{visualization}} condition, the task was the same and participants could use query languages and interactive visualization for the task. Once participants finished the task, we asked them if they noticed any patterns that might have made a CV model struggle on those images. After the completion of the task, we followed up with a survey and interview that asked about their user experience.


%% file: sections/05-results.tex
\section{Results}

\begin{table*}[h!]
\centering
\begin{tabular}{||c c c c||} 
 \hline
  & \shortstack{Baseline \\ median, mean of ranks} & \shortstack{Visualization \\ median, mean of ranks} & Mann-Whitney U test \\ [0.5ex] 
 \hline\hline
 \textbf{*Mental} & \textbf{4,13.2} & \textbf{3,7.8} & \textbf{*U = 23, X = 2, P ${<}$ 0.05, r = 0.45 (medium effect)} \\ 
 \hline
 Physical & 1.5,10 & 1.5,11 & U = 45, X = -0.34, P ${>}$ 0.05, r = 0.07 (low effect) \\
 \hline
 Temporal & 3, 11.65 & 2.5, 9.35 & U = 38.5, X = 0.83, P ${>}$ 0.05, r = 0.18 (low effect) \\
 \hline
 \textbf{*Effort} & \textbf{3, 13.3} & \textbf{2.5, 7.7} & \textbf{*U = 22, X = 2.08, P ${<}$ 0.05, r = 0.45 (medium effect)} \\
 \hline
 Stress & 2, 12.1 & 1, 8.9 & U = 34, X = 1.17, P ${>}$ 0.05, r = 0.26 (low effect) \\ [1ex] 
 \hline
\end{tabular}
\caption{Summary of participants' reported subjective workload. * denotes significance.}
\label{table:workload}
 \end{table*}


\textbf{User Performance: Identify Images, Where CV Model Struggles.}
 We asked participants to capture images where the CV models in \app~made prediction errors. We categorized the reasons participants hypothesized a model made an error into 11 error patterns (e.g., images without color (grayscale) or motion blur) using thematic analysis, which can be applied to classification and detection models. For our user performance analysis, we tracked the total number of images participants captured for each error pattern, and how many people captured each error pattern.



For the \textbf{\textit{classification}} task, participants captured images per error pattern on average was higher in the \textit{visualization} (M = 20.27, SD = 16.14) than in the \textit{baseline} condition (M = 6.36, SD = 7.59). Also, participants captured images during the task that led to a particular error pattern on average was higher in the \textit{visualization} (M = 5.63, SD = 2.46) than in the \textit{baseline} condition (M = 3.63, SD = 2.5). For the \textbf{\textit{detection}} task, participants captured images per error pattern on average was higher in the 
\textit{visualization} condition (M = 32.45, SD = 32.81) than in the \textit{baseline} (M = 14.45, SD = 15.37).
Also, participants captured images during the task that led to a particular error pattern on average was higher in the \textit{visualization} (M = 6.36, SD = 3.13) than in the \textit{baseline} condition (M = 4.27, SD = 2.49).

 Our results showed that participants in \textit{visualization} condition found more images that contained prediction errors and more variety of error patterns for both \textit{classification} and \textit{detection} tasks. Additionally, the total number of images captured per error was higher in the \textit{visualization} condition. For example, participants in \emph{visualization} condition found scarcer error patterns (e.g., scenes that were less common in the video or only a few participants identified) that participants in \emph{baseline} condition did not find. This may be because the scatter plot view helped participants find corner cases, where the CV model struggled. Participants in the \emph{visualization} condition found more easier to find error patterns (e.g., low light level, noisy background). These error patterns were easier to find because these scenes were common in the ingested videos. We hypothesize that this is because the timeline view assisted them in capturing similar images that exist within a time window. Similar images likely appear for a certain duration of time in a time-series view, which helped participants to sample those images quickly. This can help sample images that can later be used as new labels to potentially improve the current CV model.


\noindent
\textbf{Survey Results: Usability \& Subjective workload.}
We measured subjective workload in performing user tasks and usability of two conditions. To analyze the data, we used Mann-Whitney's U test (a non-parametric version of the t-test).

\textbf{Subjective workload.} We asked participants to self-report 5 subjective workloads using a Likert scale (1-very low, 5-very high)~\cite{hart2006nasa}, which were mental demand, physical demand, temporal demand, effort level, and stress level.  See Table~\ref{table:workload} for a summary.

\textbf{Usability.} We collected SUS scores~\cite{brooke1996quick} and our results showed that participants in \emph{visualization} (median = 77.25 (good)) condition reported significantly higher usability scores than in the \emph{baseline} (median = 70.25 (okay)) condition. Mann-Whitney U test showed that the mean of ranks for \emph{baseline} was 7.15 and \emph{visualization} was 13.85; U = 16.5, X = -2.49, P ${<}$ 0.05, r = 0.55 (large effect).

%% file: sections/06-discussion.tex
\section{Discussion}

We studied how two interactive visualizations can provide improved support for two fundamental activities (planning and evaluating) within the interactive machine teaching cycle when creating CV models, in the context of \app, a system that allows subject matter experts to build classifiers and extractors for images originating from videos. We assessed these techniques against the \textit{baseline} condition and found that, first, our visualizations helped participants to better leverage and compare existing CV models to find images with prediction errors. Second, our visualizations helped participants to more quickly and easily discover prediction patterns and errors across a video's timeline. Finally, participants under \textit{visualization} condition, found more images across a wider set of potential types of prediction errors and reported significantly higher usability scores and significantly lower mental demand and effort levels in performing user tasks. These results suggest that our interactive visualizations provide useful support to the IML process of building CV models, by facilitating evaluation and finding the images that contain diverse prediction errors, which can be used to guide what next actions to take to continuously improve a model's performance. We discuss the following insights:

\textbf{To evaluate a model, use it with and compare it to other models.} We hypothesize that using and comparing multiple models can help people find useful samples to improve a CV model's performance. We observed that having access to these visualizations of a model's prediction results led to a better, more diverse set of examples in which the models struggled. Finding these images can directly lead to improving a model (e.g., labeling and potential feature selection). This can assist people in the loop, in the planning stage, to familiarize themselves with the existing dataset and assist them to apply insight and foresight~\cite{ramos2020imt,teso2022leveraging} to make the model-in-training better. Also, the ability to use and compare different models opens the door to the notion of what we are calling helper concepts. A helper concept involves using a simple model, using only a small sample as a microcosm representative of the larger sample, and applying the lessons derived from that microcosm to the whole.

\textbf{Support people looking at the data globally and locally.} Our observations suggest that our visualizations can lead to insights about a model's performance that go beyond inspecting a single prediction result. By comparing the outcome of the different conditions in our study, we see that participants using interactive visualizations can better assess a model's prediction patterns globally and locally. For example, we observed that the \emph{scatterplot} view not only allowed participants to see clusters that suggested sets where different models have different opinions about the data but also {allowed participants} to see concentrations of data around prediction values that {suggested} performance issues with a model. These enhancements can facilitate image exploration and provide a guided search to improve CV models~\cite{chau2011apolo}.

%% file: sections/07-conclusion.tex
\section{Conclusion and future work}
The goal of our work has been to improve the experience of users involved in interactive ML and teaching processes of CV models working with images originating from video. We have worked towards this goal by focusing on improving the evaluation and sampling stages of the training loop using interactive visualizations. While our work's contributions and insights exist in the context of a particular type of media--task and application--we suggest our work could be generalized, using helper models to assist in sampling engineering tasks beyond CV. Also, our insights into providing ways for end-users to inquire about model behavior locally, globally, and across time can be advantageous in other scenarios involving data with a time component, such as sensor signals.

%% file: template.bbl
\begin{thebibliography}{10}

\bibitem{adadi2018peeking}
A.~Adadi and M.~Berrada.
\newblock Peeking inside the black-box: a survey on explainable artificial intelligence (xai).
\newblock {\em IEEE access}, 6:52138--52160, 2018.

\bibitem{amershi2014power}
S.~Amershi, M.~Cakmak, W.~B. Knox, and T.~Kulesza.
\newblock Power to the people: The role of humans in interactive machine learning.
\newblock {\em Ai Magazine}, 35(4):105--120, 2014.

\bibitem{bochkovskiy2020yolov4}
A.~Bochkovskiy, C.-Y. Wang, and H.-Y.~M. Liao.
\newblock Yolov4: Optimal speed and accuracy of object detection, 2020.

\bibitem{brooke1996quick}
J.~Brooke.
\newblock {\em "SUS-A quick and dirty usability scale." Usability evaluation in industry}.
\newblock CRC Press, June 1996.
\newblock ISBN: 9780748404605.

\bibitem{chau2011apolo}
D.~H. Chau, A.~Kittur, J.~I. Hong, and C.~Faloutsos.
\newblock Apolo: making sense of large network data by combining rich user interaction and machine learning.
\newblock In {\em Proceedings of the SIGCHI conference on human factors in computing systems}, pp. 167--176, 2011.

\bibitem{chen2016diagnostic}
D.~Chen, R.~K. Bellamy, P.~K. Malkin, and T.~Erickson.
\newblock Diagnostic visualization for non-expert machine learning practitioners: A design study.
\newblock In {\em 2016 IEEE Symposium on Visual Languages and Human-Centric Computing (VL/HCC)}, pp. 87--95. IEEE, 2016.

\bibitem{choo2018visual}
J.~Choo and S.~Liu.
\newblock Visual analytics for explainable deep learning.
\newblock {\em IEEE computer graphics and applications}, 38(4):84--92, 2018.

\bibitem{dang2022ganslider}
H.~Dang, L.~Mecke, and D.~Buschek.
\newblock Ganslider: How users control generative models for images using multiple sliders with and without feedforward information.
\newblock In {\em CHI Conference on Human Factors in Computing Systems}, pp. 1--15, 2022.

\bibitem{dudley2018review}
J.~J. Dudley and P.~O. Kristensson.
\newblock A review of user interface design for interactive machine learning.
\newblock {\em ACM Transactions on Interactive Intelligent Systems (TiiS)}, 8(2):1--37, 2018.

\bibitem{fails2003interactive}
J.~A. Fails and D.~R. Olsen~Jr.
\newblock Interactive machine learning.
\newblock In {\em Proceedings of the 8th international conference on Intelligent user interfaces}, pp. 39--45, 2003.

\bibitem{fiebrink2010wekinator}
R.~Fiebrink and P.~R. Cook.
\newblock The wekinator: a system for real-time, interactive machine learning in music.
\newblock In {\em Proceedings of The Eleventh International Society for Music Information Retrieval Conference (ISMIR 2010)(Utrecht)}, vol.~3, 2010.

\bibitem{gou2020vatld}
L.~Gou, L.~Zou, N.~Li, M.~Hofmann, A.~K. Shekar, A.~Wendt, and L.~Ren.
\newblock Vatld: a visual analytics system to assess, understand and improve traffic light detection.
\newblock {\em IEEE transactions on visualization and computer graphics}, 27(2):261--271, 2020.

\bibitem{hart2006nasa}
S.~G. Hart.
\newblock Nasa-task load index (nasa-tlx); 20 years later.
\newblock In {\em Proceedings of the human factors and ergonomics society annual meeting}, vol.~50, pp. 904--908. Sage publications Sage CA: Los Angeles, CA, 2006.

\bibitem{he2015deep}
K.~He, X.~Zhang, S.~Ren, and J.~Sun.
\newblock Deep residual learning for image recognition, 2015.

\bibitem{hohman2020understanding}
F.~Hohman, K.~Wongsuphasawat, M.~B. Kery, and K.~Patel.
\newblock Understanding and visualizing data iteration in machine learning.
\newblock In {\em Proceedings of the 2020 CHI conference on human factors in computing systems}, pp. 1--13, 2020.

\bibitem{holzinger2019interactive}
A.~Holzinger, M.~Plass, M.~Kickmeier-Rust, K.~Holzinger, G.~C. Cri{\c{s}}an, C.-M. Pintea, and V.~Palade.
\newblock Interactive machine learning: experimental evidence for the human in the algorithmic loop.
\newblock {\em Applied Intelligence}, 49(7):2401--2414, 2019.

\bibitem{kraska2018northstar}
T.~Kraska.
\newblock Northstar: An interactive data science system.
\newblock {\em Proceedings of the VLDB Endowment}, 11(12):2150--2164, 2018.

\bibitem{liu2017towards}
S.~Liu, X.~Wang, M.~Liu, and J.~Zhu.
\newblock Towards better analysis of machine learning models: A visual analytics perspective.
\newblock {\em Visual Informatics}, 1(1):48--56, 2017.

\bibitem{cosmosdb-docs}
Microsoft.
\newblock Getting started with sql queries, 2021.

\bibitem{customvision2019}
{Microsoft}.
\newblock What is azure custom vision?, 2021.

\bibitem{ng2020understanding}
F.~Ng, J.~Suh, and G.~Ramos.
\newblock Understanding and supporting knowledge decomposition for machine teaching.
\newblock In {\em ACM conference on Designing Interactive Systems (DIS)}, July 2020.

\bibitem{oquab2014learning}
M.~Oquab, L.~Bottou, I.~Laptev, and J.~Sivic.
\newblock Learning and transferring mid-level image representations using convolutional neural networks.
\newblock In {\em Proceedings of the IEEE conference on computer vision and pattern recognition}, pp. 1717--1724, 2014.

\bibitem{ramos2020imt}
G.~Ramos, C.~Meek, P.~Simard, J.~Suh, and S.~Ghorashi.
\newblock Interactive machine teaching: a human-centered approach to building machine-learned models.
\newblock {\em Human–Computer Interaction}, 35(5-6):413--451, 2020. doi: {{%
10\hspace{.1pt}\discretionary{.}{%
}{.}\hspace{.4pt}1080\discretionary{/}{%
}{/}07370024\hspace{.1pt}\discretionary{.}{%
}{.}\hspace{.4pt}2020\hspace{.1pt}\discretionary{.}{%
}{.}\hspace{.4pt}1734931}}


\bibitem{ren2016squares}
D.~Ren, S.~Amershi, B.~Lee, J.~Suh, and J.~D. Williams.
\newblock Squares: Supporting interactive performance analysis for multiclass classifiers.
\newblock {\em IEEE transactions on visualization and computer graphics}, 23(1):61--70, 2016.

\bibitem{ribeiro2016should}
M.~T. Ribeiro, S.~Singh, and C.~Guestrin.
\newblock " why should i trust you?" explaining the predictions of any classifier.
\newblock In {\em Proceedings of the 22nd ACM SIGKDD international conference on knowledge discovery and data mining}, pp. 1135--1144, 2016.

\bibitem{simard2017machine}
P.~Y. Simard, S.~Amershi, D.~M. Chickering, A.~E. Pelton, S.~Ghorashi, C.~Meek, G.~Ramos, J.~Suh, J.~Verwey, M.~Wang, et~al.
\newblock Machine teaching: A new paradigm for building machine learning systems.
\newblock {\em arXiv preprint arXiv:1707.06742}, 2017.

\bibitem{steinfeld2009oz}
A.~Steinfeld, O.~C. Jenkins, and B.~Scassellati.
\newblock The oz of wizard: simulating the human for interaction research.
\newblock In {\em Proceedings of the 4th ACM/IEEE international conference on Human robot interaction}, pp. 101--108, 2009.

\bibitem{suh2019anchorviz}
J.~Suh, S.~Ghorashi, G.~Ramos, N.-C. Chen, S.~Drucker, J.~Verwey, and P.~Simard.
\newblock Anchorviz: Facilitating semantic data exploration and concept discovery for interactive machine learning.
\newblock {\em ACM Transactions on Interactive Intelligent Systems (TiiS)}, 10(1):1--38, 2019.

\bibitem{sultanum2020teaching}
N.~Sultanum, S.~Ghorashi, C.~Meek, and G.~Ramos.
\newblock A teaching language for building object detection models.
\newblock In {\em Proceedings of the 2020 ACM Designing Interactive Systems Conference}, pp. 1223--1234, 2020.

\bibitem{teso2022leveraging}
S.~Teso, {\"O}.~Alkan, W.~Stammer, and E.~Daly.
\newblock Leveraging explanations in interactive machine learning: An overview.
\newblock {\em arXiv preprint arXiv:2207.14526}, 2022.

\bibitem{wu2022survey}
X.~Wu, L.~Xiao, Y.~Sun, J.~Zhang, T.~Ma, and L.~He.
\newblock A survey of human-in-the-loop for machine learning.
\newblock {\em Future Generation Computer Systems}, 2022.

\bibitem{yosinski2014transferable}
J.~Yosinski, J.~Clune, Y.~Bengio, and H.~Lipson.
\newblock How transferable are features in deep neural networks?
\newblock {\em arXiv preprint arXiv:1411.1792}, 2014.

\bibitem{youtube}
YouTube.
\newblock Hospital construction, 2021.

\bibitem{zhang2018visual}
Q.-s. Zhang and S.-C. Zhu.
\newblock Visual interpretability for deep learning: a survey.
\newblock {\em Frontiers of Information Technology \& Electronic Engineering}, 19(1):27--39, 2018.

\end{thebibliography}
